\documentclass{iau}
\usepackage{graphicx}

\def\etal       {et~al.}
\def\muas     {~$\mu$as}

\title[RadioAstron space-VLBI project: studies of masers] 
{RadioAstron space-VLBI project: \\ studies of masers \\ in star forming regions of our Galaxy \\ and megamasers in external galaxies }

\author[A.M. Sobolev \& RadioAstron Maser Team]   
{A.M.\,Sobolev$^1,$
N.N.\,Shakhvorostova$^2$,
A.V.\,Alakoz$^2$,
\and W.A.\,Baan$^3$  on behalf of the RadioAstron maser team}

\affiliation{$^1$Astronomical Observatory, Ural Federal University, \\ Lenin Ave. 51,
Ekaterinburg 620083, Russia \\ email: {\tt Andrej.Sobolev@urfu.ru} \\[\affilskip]
$^2$Astro-Space Center of LPI RAS, Moscow, Russia \\[\affilskip]
$^3$ASTRON, Netherlands}

\pubyear{2017}
\volume{336}  
\jname{Astrophysical Masers: Unlocking the Mysteries of the Universe}
\editors{A. Tarchi, M.J. Reid \& P. Castangia, eds.}
\begin{document}

\maketitle

\begin{abstract}
   Observations of the masers in the course of RadioAstron mission yielded detections of fringes for a number of sources in both water and hydroxyl maser transitions. Several sources display numerous ultra-compact details. This proves that implementation of the space VLBI technique for maser studies is possible technically and is not always prevented by the interstellar scattering, maser beaming and other effects related to formation, transfer, and detection of the cosmic maser emission.
   For the first time, cosmic water maser emission was detected with projected baselines exceeding Earth Diameter. It was detected in a number of star-forming regions in the Galaxy and megamaser galaxies NGC~4258 and NGC~3079.
   RadioAstron observations provided the absolute record of the angular resolution in astronomy. Fringes from the NGC~4258 megamaser were detected on baseline exceeding 25 Earth Diameters. This means that the angular resolution sufficient to measure the parallax of the water maser source in the nearby galaxy LMC was directly achieved in the cosmic maser observations. 
   Very compact features with angular sizes about 20\muas\, have been detected in star-forming regions of our Galaxy. Corresponding linear sizes are about 5-10 million kilometers.
   So, the major step from milli- to micro-arcsecond resolution in maser studies is done in the RadioAstron mission. The existence of the features with extremely small angular sizes is established. Further implementations of the space--VLBI maser instrument for studies of the nature of cosmic objects, studies of the interaction of extremely high radiation field with molecular material and studies of the matter on the line of sight are planned.

\end{abstract}

\firstsection 
\section{Introduction}

The space-ground interferometer RadioAstron allows observations with the longest-ever baselines exceeding the size of the Earth by more than an order of magnitude. Maser sources represent one of the main targets of the RadioAstron (RA) science program along with active galactic nuclei and pulsars.
The RadioAstron project allows us to observe maser emission in one quantum transition of water at 22.235 GHz and two transitions of hydroxyl at 1.665 and 1.667 GHz. Water and hydroxyl masers are found in star-forming regions of our and nearby galaxies, around mass-loosing evolved stars, and in accretion discs around super-massive black holes in external galaxies.

Masers have small angular sizes (a few milli-arcsec and smaller), very high flux densities (up to hundreds of thousand Jy), and small line widths (normally about 0.5 km/s and smaller). Because of that masers proved to be precise instruments for studies of kinematics and physical parameters of the objects in our and other galaxies.

The space radio interferometer RadioAstron provides a record of high angular resolution. This provides tight limits on the sizes of the most compact maser spots and estimates of their brightness temperatures, which are necessary input for the studies of the pumping mechanisms.

Typical values of the minimal flux density detectable with RA for the water masers at 22~GHz and hydroxyl masers at 1.665/1.667~GHz are 15~Jy and 3.5~Jy, respectively. These values were calculated for a typical line width of 0.1~km/sec and coherent accumulation time of 100~sec and 600~sec for 22~GHz and 1.665/1.667~GHz, respectively. However, when we use the large ground-based antenna, for example 100-m GBT, and the line is broad, RA proved ability to detect 3-4~Jy water maser source.

\section{Statistics of maser observations for the first 6 years of operation}

{\underline{\it Maser observation program}}. During the period from November 2011 to May 2012 interferometric mode of RA operation was tested. For that purpose, a number of bright quasars and the brightest and most compact sources of maser emission were selected~(\cite{raes}). Basic conditions for choosing these sources were the existence of details that remain compact (i.e. unresolved) on the longest baseline projections and the highest brightness temperatures measured during VLBI and VSOP surveys. The first positive detections of maser sources by space interferometer were achieved for W51 (water) and W75N (hydroxyl) in two sessions in May and July 2012. Baseline projections were 1.0$-$1.5 and 0.1$-$0.8~Earth diameters (ED), respectively. Later, more sophisticated data analysis led to even more positive results in these test sessions: compact water maser features were detected in W3~IRS5 and W3(OH) in two sessions in February 2012. Baseline projections were 3.7$-$3.9~ED.

After the first successful tests, the early science program started. The main purpose of these observations was to obtain first astrophysical results and measurements of the main parameters of the operating interferometer. List of observed sources was significantly expanded, objects of other types were included in addition to the star-forming regions. Stellar masers in S~Per, VY~CMa, NML~Cyg, U~Her and extragalactic masers in Circinus and N113 were observed. It was proved that RadioAstron can observe cosmic masers with very high spectral resolution. This was not obvious at the beginning, indicates the presence of the ultra-fine structure in the maser images, and that interstellar scattering does not prevent observations of masers in the galactic plane (\cite{raes}). Positive detections for stellar and extragalactic masers were not obtained during the early science program.

The early science program was followed by the key and general research programs which were conducted (and the general program continues at the moment) on the basis of the open call for proposals received from research teams around the world. Details of the preparation and the conditions of the call for proposals are published at the RadioAstron project site (\textit{http://www.asc.rssi.ru/radioastron}). The main objectives of this phase of the maser program are studying the kinematics and dynamics of the compact sources of maser emission in star-forming regions, as well as the study of extragalactic masers. As a result, the signals from extragalactic masers NGC3079 and NGC~4258 were detected along with star-forming regions. The maser of NGC~4258 is associated with the accretion disk around super-massive black hole at the center of this galaxy. Projected baselines in these observing sessions were up to 2.0 ED (for NGC3079) and exceeding 25 ED (for NGC~4258) corresponding to angular resolutions of 115~$\mu$as and 8~$\mu$as, respectively. The latter resolution represents a record of angular resolution in astronomy, the previous record of 11~$\mu$as was reported by us in (\cite{ra_rec}). The new record resolution is formally sufficient to measure the parallax of the water maser source in the nearby galaxy LMC.

\begin{table}
\caption{Maser sources detected on the space--ground baselines}
\centering
\begin{tabular}{|c|c|c|}
\hline
\bf{Source} &  \bf{Projected baseline length, ED} & \bf{Best angular} \\
 & & \bf {resolution, $\mu$as} \\
\hline
\bf{Galactic} & & \\
\bf{H$_2$O masers} & \bf{0.4$-$10.0} & \bf{22} \\
\hline
W3 Irs5 & 2.5$-$2.8; 3.5; 3.9; 5.4; &  \\
        & 6.0; 6.0$-$10.0 & 22 \\
\hline
W49 N & 2.2$-$3.0; 4.5; 7.9; 9.4 & 23 \\
\hline
W3(OH) & 3.9 & 56 \\
\hline
Cepheus A & 0.9$-$1.7; 1.1; 3.1$-$3.5 & 62 \\
\hline
Orion KL & 1.9; 3.4 & 64 \\
\hline
W51 E8 & 0.4$-$2.3; 1.3; 1.4$-$1.8; 1.7 & 95 \\
\hline
G43.8$-$00.13 & 1.2; 1.2 & 182 \\
\hline
\bf{Extragalactic} & & \\
\bf{H$_2$O masers} & \bf{1.3$-$greater 25} & \bf{less than 9} \\
\hline
NGC~4258 & 1.3; 1.7; 6.5; 9.5; 11.6; 11.8;  &  \\
         & 12.2; 19.2; 19.5, greater 25  & less than 9 \\
\hline
NGC~3079 & 1.9; 1.9 & 115 \\
\hline
\bf{Galactic} &  &  \\
\bf{OH masers} & \bf{0.1$-$1.9} & \bf{1540} \\
\hline
Onsala 1 & 0.2$-$0.7; 1.0$-$1.9 & 1540 \\
\hline
W75 N & 0.1$-$0.3; 0.1$-$0.8 & 3660 \\
\hline
\end{tabular}
\label{detect_sources}
\end{table}


{\underline{\it General statistics of maser source detections}}. This section provides statistics from the beginning of RA maser observations (Nov 2011) up to present time (Oct 2017). During this period a large amount of data was accumulated. 154 maser observation sessions were conducted, and 32 sources were observed. The majority of masers observed in RA program is related to star-forming regions~-- 20 sources in total. 8 maser sources in the envelopes of late-type stars of the Galaxy were observed, and 4 extragalactic masers in star-forming regions and circum-nuclear disks of external galaxies were observed.

The scientific data have been corrupted or lost in 10 sessions out of total 154 due to technical problems. 141 observations of the remaining 144 sessions at the moment (Oct 2017) are processed on the ASC software correlator (\cite{corr}), positive detections are obtained in 38 sessions. Thus, the current detection rate of fringes at space-ground baselines is about 27~\%.

All of the successful fringe detections for galactic masers at space-ground baselines were obtained for the sources associated with star-forming regions,~-- 26 of all 38 positive detections. 12 detections were obtained for extragalactic masers. No fringes for the stellar masers were obtained at space-ground baselines yet.

\begin{figure}[t]
\begin{center}
 \includegraphics[width=12.0cm]{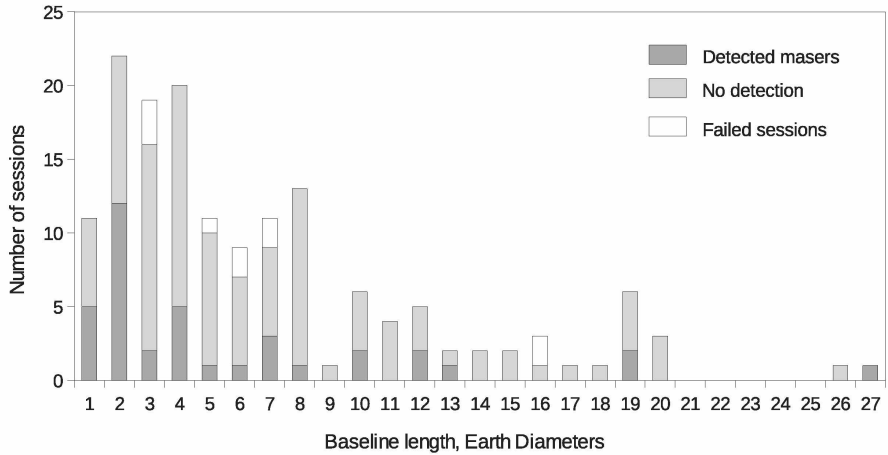}
 \caption{Statistics of maser observation results over projected baseline length of the space interferometer RadioAstron.}
   \label{fig1}
\end{center}
\end{figure}

Table ~\ref{detect_sources} gives information on the observational sessions which provided positive fringe detection with the space-ground interferometer. The columns show source names, projected lengths of the space-ground baselines at which the interferometric detection was obtained and the last column shows the best angular resolution achieved for each source. Each baseline (or baseline interval) corresponds to one observational session with positive detection.

The distribution of the number of detections depending on the length of the baseline projection is shown in the Figure~1, which presents statistics of observational sessions for the whole set of database lengths. It is seen that most of the positive detections fall in the range from 1 to 4 ED.

\section{Summary}
\label{results}

The main conclusions of the work are the following:

1. Space-VLBI observations of the water and hydroxyl masers show that the bright details of the masers in galactic star-forming regions often remain unresolved at projected baseline which exceeds Earth diameter many times.

2. Record resolution better than 9 \muas\, for any astronomical observation is obtained for NGC~4258 extragalactic water maser. Scintillation does not prevent fringe detection.

3. Very compact water maser features with the angular sizes of about 20$-$60 \muas\, are registered in galactic star-forming regions. Their brightness temperatures range from $10^{14}$ up to $10^{16}$~K (\cite{Shakh17}). The best linear resolution better than 4 million km (a few solar diameters) was achieved for Orion maser. The best for galactic maser angular resolution of 23 micro-arcsec was achieved for W~49~N. This source is located in the galactic plane about 11 kpc away. So, detection provides important input for the theory of interstellar scattering in the Galaxy.

So, the major step from milli- to micro-arcsecond resolution in maser studies is done in the RadioAstron mission. The existence of the features with extremely small angular sizes is established. Further implementations of the space--VLBI maser instrument for studies of the nature of cosmic objects, studies of the interaction of extremely high radiation field with molecular material and studies of the matter on the line of sight are planned.

{\underline{\it Acknowledgements}}. The RadioAstron project is led by the Astro Space Center of the Lebedev Physical Institute of the Russian Academy of Sciences and the Lavochkin Association of the Russian Federal Space Agency, and is a collaboration with partner institutions in Russia and other countries. This research is partly based on observations with the 100 m telescope of the MPIfR at Effelsberg; radio telescopes of IAA RAS (Institute of Applied Astronomy of Russian Academy of Sciences); Medicina \& Noto telescopes operated by INAF; Hartebeesthoek, Torun, WSRT, Yebes, and Robledo radio observatories. The National Radio Astronomy Observatory is a facility of the National Science Foundation operated under cooperative agreement by Associated Universities, Inc. Results of optical positioning measurements of the Spektr-R spacecraft by the global MASTER Robotic Net, ISON collaboration, and Kourovka observatory were used for spacecraft orbit determination in addition to mission facilities. AMS was financially supported by the Russian Science Foundation (project no. 15-12-10017).

\end{document}